\documentclass[%
 nofootinbib,
 aip,
 jmp,%
 amsmath,amssymb,
 reprint,%
]{revtex4-1}

\usepackage{graphicx}
\usepackage{dcolumn}
\usepackage{bm}

\begin{document}


\title[World sheets of spinning particles]{World sheets of spinning particles}
\author{D.~S.~Kaparulin}
\email{dsc@phys.tsu.ru}
\author{S.~L.~Lyakhovich}%
 \email{sll@phys.tsu.ru}
\affiliation{ Physics Faculty, Tomsk State University, Tomsk 634050,
Russia}%

\date{\today}

\begin{abstract}
The classical spinning particles are considered such that
quantization of classical model leads to an irreducible massive
representation of the Poincar\'e group. The class of gauge
equivalent classical particle world lines is shown to form a
\mbox{$[(d+1)/2]$}-dimensional world sheet in $d$-dimensional
Minkowski space, irrespectively to any specifics of the classical
model. For massive spinning particles in $d=3,4$, the world sheets
are shown to be circular cylinders. The radius of cylinder is fixed
by representation. In higher dimensions, particle's world sheet
turns out to be a toroidal cylinder $\mathbb{R}\times \mathbb{T}^D$,
$D=[(d-1)/2]$. Proceeding from the fact that the world lines of
irreducible classical spinning particles are cylindrical curves,
while all the lines are gauge equivalent on the same world sheet, we
suggest a method to deduce the classical equations of motion for
particles and also to find their gauge symmetries. In $d=3$
Minkowski space, the spinning particle path is defined by a single
fourth-order differential equation having two zero-order gauge
symmetries. The equation defines particle's path in Minkowski space,
and it does not involve auxiliary variables. A special case is also
considered of cylindric null curves, which are defined by a
different system of equations. It is shown that the cylindric null
curves also correspond to irreducible massive spinning particles.
For the higher-derivative equation of motion of the irreducible
massive spinning particle, we deduce the equivalent second-order
formulation involving an auxiliary variable. The second-order
formulation agrees with a previously known spinning particle model.
\end{abstract}

\maketitle

\section{Introduction}
\noindent The history of classical spinning particle models has
developed over nine decades starting from the work by Frenkel
\cite{Fren}. For a brief summary of the first seventy years and
corresponding bibliography we refer to review \cite{Fryd}. In the
introduction, we discuss some generalities about irreducible
classical spinning particle models, and then explain the idea of
describing the particle dynamics by the world sheets of certain
type.

The common feature shared by all the classical spinning particle
models is that their degrees of freedom include, besides particle's
position in space, also the position in some ``internal space.'' In
geometric terms, the configuration space $\mathcal{M}$ of spinning
particle is a fiber bundle over Minkowski space, with coordinates on
fibers describing configurations of spinning degrees of freedom.
$\mathbb{Z}_2$-grading can be assigned to the fibers, so the
spinning degrees of freedom can be described by Grassmann-odd or
-even coordinates. Be the fibers even or odd, they can describe both
integer and half-integer spins. Once the Poincar\'e-invariant action
functional is found for the trajectories on $\mathcal{M}$, the
particle model is defined at classical level, and it can be
quantized.

Once the classical model has Poincar\'e symmetry, the space of
quantum states should carry the Poincar\'e group representation
which can be reducible, or irreducible. The
Kirillov-Konstant-Soureau method \cite{Kirillov,Kostant,Sour} tells
us that if the quantum mechanical system realizes irreducible
representation, the classical limit is the dynamical system
corresponding to the coadjoint orbit of the group. The classical
action functional of the system is defined by symplectic form on the
co-orbit. In this way, we know that the physical phase space of the
classical irreducible massive spinning particle is a coadjoint orbit
of the Poincar\'e group corresponding to nonvanishing mass and spin.
The latter fact means \cite{LSS} that for any irreducible classical
spinning particle model, irrespectively to any specifics of fibers
over Minkowski space chosen to describe spinning degrees of freedom,
all the on-shell gauge invariant observables have to be functions of
the conserved momentum $p$, and angular momentum $J$. The admissible
values of $p$ and $J$ are restricted only by the spin-shell and
mass-shell constraints.

Given the above-mentioned understanding of irreducibility at
classical level, the spinning particle models are typically
constructed in three main stages. First, the particle's
configuration space $\mathcal{M}$ is chosen to be a fiber bundle
over Minkowski space. This bundle is given by the Cartesian product
$\mathcal{M}=\mathbb{R}^{1,d-1}\times S$ of Minkowski space to some
manifold $S$ (the typical fiber). The manifold $S$ is supposed to be
a homogeneous space of the Poincar\'e group. In the old-fashioned
terminology, the models are categorized by types of coordinates
introduced in the internal space $S$. In this way, the vectorial,
tensorial, spinorial, twistorial, etc. types of spinning particle
models are distinguished \cite{Fryd}. At the second stage, all the
Poincar\'e invariants are identified of the trajectories on
$\mathcal{M}$ (usually without higher derivatives), and the most
general ansatz is constructed for the action functional. At the
third stage, the action is specified in the way that makes the model
irreducible. The latter means that all the on-shell gauge invariants
are functions of Noether's momentum $p$ and angular momentum $J$.
These conserved quantities must satisfy the mass-shell and
spin-shell conditions, being free from any other constraint.

In works \cite{LSS,KuLS1}, the configuration space of the spinning
particle has been suggested to be the product of Minkowski space and
2-sphere, $\mathcal{M}_6=\mathbb{R}^{1,3}\times S^2$. These models
can be viewed as minimal in the sense that $S^2$ is a manifold of
the minimal possible dimension where the Poincar\'e group acts
transitively. In the paper \cite{KuLS1}, two Poincar\'e invariants
were found of the trajectories on $\mathcal{M}_6$, while in paper
\cite{LSS} one more Poincar\'e invariant was found, being of
Wess-Zumino type. Once all the invariants of trajectories on
$\mathcal{M}_6$ are included in the action, the model becomes
universal. Depending on the constant parameters in the Lagrangian,
the model is able to describe the irreducible massive and massless
spinning particle, and irreducible particle with continuous
helicity. In the massive case, with all the invariants involved in a
special way, the model admits consistent inclusion of interactions
with general electromagnetic field and gravity. The model without
third invariant \cite{KuLS1} describes the irreducible massive
spinning particle, though it admits inclusion of interaction only
with the constant curvature gravity \cite{KuLSS1} and constant
electromagnetic field. The minimal models of \cite{LSS,KuLS1}, were
rederived for various reasons, reformulated, and reinterpreted in
various ways, see
\cite{Starus,Bratek1,Bratek2,Bratek3,DasGhosh,Kassaetall,Rempel} and
references therein. The most frequent common feature of the
reformulations is that some linear space $\mathcal{L}$ of higher
dimension is considered as the phase space of spinning particle
instead of $T^*\mathcal{M}_6$, while some extra constraints are
imposed on linear coordinates. Most typically, the constraints are
quadratic in coordinates. They reduce $\mathcal{L}$ to
$T^*\mathcal{M}_6$. For example, the ``bilocal'' irreducible
spinning particle model was recently developed in Ref. \cite{Rempel}
with the phase space being a squared cotangent bundle to Minkowski
space, $\mathcal{L}=T^*\mathbb{R}^{1,3}\times T^*\mathbb{R}^{1,3}$.
The quadratic constraints are imposed on $\mathcal{L}$ reducing the
dynamics to sub-bundle $T^*\mathcal{M}_6\subset\mathcal{L}$, and the
model turns out equivalent to that proposed in \cite{KuLS1}. The
minimal models of irreducible spinning particles are also known for
higher dimensions \cite{LSS6d,LSShd1,LSShd1/2} and in $d=3$
\cite{GKL1,AL}. In the latter case, the configuration space is
$\mathcal{M}_4=\mathbb{R}^{1,2}\times S^1$.

A somewhat special class of spinning particle models are
higher-derivative theories. No internal  space is engaged, while the
spinning degrees of freedom come from extra derivatives involved in
the equations of motion in Minkowski space. The first model with a
Lagrangian involving the arc length and curvature of particle's
world line was suggested by Pisarski \cite{Pisar}. Later, the
particle models with higher-derivative invariants included in the
Lagrangian were studied from various viewpoints. We mention here
only some of these works which particularly address the issue of
irreducibility of the models. Once the curvature and arc length are
both included linearly in the action, the model is reducible with
the certain Regge trajectory connecting mass and spin
\cite{Ply-rigid}. It was also noticed that the linear in curvature
Lagrangian has an extra gauge symmetry of W-type and describes the
irreducible massless spinning particle \cite{Plyu-massless,RR1,RR2}.
These conclusions extend to a more general action and to higher
dimensions: linear in curvature actions describe irreducible
massless spinning particles, while any other combination of the arc
length, and/or higher-derivative invariants results in reducible
representations \cite{Ners1,Ners2}. It is also relevant to mention
the model with the Lagrangian, being the curvature, with the
trajectories restricted to the class of null curves
\cite{NerRam1,NerRam2}.\footnote{In $d=3$, the more general
Lagrangian, being a linear combination of the curvature and torsion,
has been considered in \cite{NerMan}.} It corresponds to a reducible
representation involving massive states \cite{NerRam1,NerRam2}. The
attractive feature of higher-derivative spinning particle models is
that they describe the dynamics of point particles in terms of lines
in Minkowski space, without a recourse to any internal space. The
irreducible models are known in terms of Minkowski space
trajectories only for massless particles, while all the known
higher-derivative models for massive particles are reducible. In
this article, we show that the irreducible massive spinning particle
dynamics can be described by higher-derivative equations for its
paths in Minkowski space, without introducing any auxiliary
configuration space to accommodate the spinning degrees of freedom.

Let us announce the key observation we proceed from: once the
massive spinning particle  is irreducible, the equivalence class of
classical particle paths in $d$-dimensional Minkowski space forms a
$[(d+1)/2]$-dimensional surface. By equivalent paths we understand
the trajectories which are connected by gauge transformations. In
other words, the trajectories of irreducible massive spinning
particle are world sheets rather than world lines. The space of all
world sheets in Minkowski space turns out to be isomorphic to the
Poincar\'e group co-orbit for the corresponding representation. The
geometry of the spinning particle world sheets turns out very
simple. In $d=3$ and $d=4$, the world sheets are $2d$ cylinders of
fixed radius with timelike axis. The cylinder axis is directed along
the particle's momentum, while the position of the axis in space is
determined by the particle's angular momentum. The radius of
cylinder is fixed by spin. In higher dimensions, the particle world
sheets are the toroidal cylinders, $\mathbb{R}\times
\mathbb{T}^{[(d-1)/2]}$, where $\mathbb{T}^{[(d-1)/2]}$ is
$[(d-1)/2]$-torus. The radii of the torus are fixed by the
eigenvalues of the Casimir operators corresponding to an irreducible
representation.

There is a subtle issue related to the fact that different world
sheets can intersect with each other, so some special world lines
can belong to several world sheets. We will term the world lines
atypical paths if they belong to different world sheets. Once the
world line belongs to a unique world sheet, it is termed a typical
path. Given the typical path, it defines the world sheet, while the
atypical one does not. As we shall see, all the typical paths are
gauge equivalent on the same world sheet, while the atypical paths
do not mix up by gauge transformations with any other line. As the
physical phase space of the irreducible particle, being understood
as the Poincar\'e group co-orbit, is isomorphic to the variety of
the particle world sheets, we consider only typical world lines as
equivalence class of observable paths. The atypical lines are
excluded, because they do not represent a unique world sheet.

In minimal models of irreducible spinning particles in various
dimensions, it has been previously noticed \cite{KuLS1,LSS6d,GKL1}
that particle's paths in Minkowski space always lie on a surface
with the symmetry axis being the conserved momentum. This is
interpreted as the \emph{Zitterbewegung} phenomenon. A somewhat
similar phenomenon occurs in some massless spinning particle models.
In particular, the paths of chiral fermions in $d=4$  are noticed to
lie on $3$-planes \cite{Duval1,Duval2}. What we state here goes
beyond these observations in several respects. The first is that the
very fact of irreducibility of massive spinning particle constrains
particle's paths in Minkowski space to be cylindric lines. This fact
does not depend on any specifics of the particle model, like the
choice of internal configuration space. Second, every causal typical
path on a specific cylinder is gauge equivalent to any other causal
line on the same cylinder\footnote{If zero component of momentum is
positive, the path is considered causal once $\dot{x}{}^0>0$. }.
Third, the definition of the particle trajectory as a cylindric line
in Minkowski space has differential consequences. These relations
between the derivatives of the path can be viewed as equations of
motion for the particle, and also as an alternative definition of
cylindric lines. These higher-derivative equations for the paths in
Minkwoski space are not Lagrangian. In $d=3$, these equations turn
out to be equivalent to Lagrangian equations of lower order
constructed for the minimal model of spinning particle \cite{GKL1}.
For higher dimensions, similar equivalence is expected, though it is
a more technically complex issue which we do not address in this
article.

\section{Irreducibility of representation and particle world sheet in Minkowski space}
\noindent Consider quantum mechanics of particle in Minkowski
space.\footnote{ We use a mostly positive signature of the Minkowski
metric throughout the paper.} The action of the Poincar\'e group in
the space of physical states is generated by the operators of
momentum $\hat{p}_a$ and angular momentum $\hat{J}_{ab}$. The
representation is implied to be unitary of the Poincar\'e group, so
the operators $\hat{p}_a, \hat{J}_{ab}$ have to be Hermitian. Once
the particle is a point object in Minkowski space, the space of
states is supposed to admit representation by functions of the
particle position $x$ and maybe some other variables of a spinning
sector. The momentum is supposed to generate translations in
Minkowski space, so
\begin{equation}\label{hatp}
\hat{p}_a x^b= i\delta_a^b \, .
\end{equation}
The angular momentum is supposed to generate Lorentz transformations
of Minkowski space, so it can be represented as a sum of orbital
momentum and spin momentum,
\begin{equation}\label{J-S}
    \hat{J}_{ab}= x_a \hat{p}_b -x_b \hat{p}_a  + \hat{S}_{ab}\, .
\end{equation}
Once Lorentz transformations of Minkowski space coordinates are
generated by the orbital momentum $x_a \hat{p}_b -x_b \hat{p}_a$,
the action of the spin generator on $x$ should be trivial,
\begin{equation}\label{Sx}
    \hat{S}_{ab}f(x)=0 \,,\qquad \forall f\in \mathcal{C}^\infty(R^{(1,d-1)})  .
\end{equation}
The latter fact also means that $[\hat{p},\hat{S}]=0$. Rel.
(\ref{J-S}) can be considered as a definition for the operator of
spin $\hat{S}$ in terms of momentum $\hat{p}$, angular momentum
$\hat{J}$ and position $x$ of the point particle in Minkowski space,
\begin{equation}\label{SpJx}
    \hat{S}= \hat{J} - x\wedge \hat{p}\,.
\end{equation}
Once the representation is irreducible of the Poincar\'e group, any
linear operator acting on the space of states, and commuting with
$\hat{p}$ and $\hat{J}$, should be a multiple of unit. This applies
not only to the elements of the universal enveloping algebra of the
Poincar\'e group, but also to any polynomial in the Casimir
operators constructed from $\hat{S}$. The spin momentum operators
$\hat{S}$ generate the Lorentz group representation which should  be
irreducible in its own turn, to avoid reducibility of the Poincar\'e
group representation. This means that for a nondegenerate
representation of the Poincar\'e group generated by the operators
(\ref{hatp}) and (\ref{J-S}), the Casimir operators $C_k(\hat{p},
\hat{J}), k=0,1, \ldots, [(d-1)/2]$ of the Poincar\'e group and the
Casimir operators of the Lorentz group $C^L_l(\hat{S}), \, l=0,1,
\,\ldots \, , [d/2]-1$ should be a multiple of unit
 \begin{eqnarray}
 C_0\equiv \hat{p}^2 = -m^2\, , \qquad  C_k (\hat{p}, \hat{J})= s_k
 \,,\qquad k=1, \ldots,[(d-1)/2]\,;
\label{C-P}\end{eqnarray}
\begin{eqnarray}\label{C-L}
 C^L_0\equiv \hat{S}^2 =  \text{sign}(\varrho)\varrho^2 \, ,\qquad  C^L_l (\hat{S})= \varrho_l
 \,,\qquad l=1, \ldots [d/2]-1\,,
\end{eqnarray}
where $m, \varrho, s_k, \varrho_l$ are real numbers being the
eigenvalues of the Casimir operators. Irreducibilty of the
representation further means that any operator commuting to all the
generators $\hat{p}, \hat{J}$ is spanned by the Casimir operators
(\ref{C-P}), (\ref{C-L}),
\begin{equation}\label{C+}
[ \hat{A}, \hat{p} ]=[ \hat{A}, \hat{J} ] =0 \,  \qquad
\Leftrightarrow  \qquad\hat{A}= f(C,C^L)\,.
\end{equation}
For example,  in $d=3$, we have three operators commuting with the
generators (\ref{hatp}), (\ref{J-S}): the operator of squared spin
$\hat{S}{}^2$, and two Casimir operators of the Poincar\'e group,
mass $\hat{p}{}^2$ and spin $W=\hat{p}\cdot \hat{J}$. In $d=4$ we
have four commuting operators: the operator of squared spin
$\hat{S}{}^2$, contraction of spin and its dual
$\frac{1}{2}\epsilon_{abcd}\hat{S}^{ab}\hat{S}^{cd}$, and two
Casimir operators of the Poincar\'e group -- mass $\hat{p}{}^2$ and
operator of squared Pauli-Lubanski vector
$W^2,W_a=\frac{1}{2}\epsilon_{abcd}\hat{J}^{bc}\hat{p}^d$. In
$d$-dimensional Minkowski space, the number of commuting operators
is $d$.

The above-mentioned obvious facts have consequences for the
classical limit of the quantum theory where the Poincar\'e group
representation is irreducible while the Lorentz generators have the
structure (\ref{J-S}). Once a quantum observable is a multiple of
unit, the corresponding classical quantity has to be constrained to
a constant. This means, in the classical limit of the irreducible
quantum theory, the constraints are imposed on $p, J, x$,
 \begin{eqnarray}
     p^2+m^2 =0\, ,\qquad C_k(p,J)=s_k\,,\qquad
k=1, \ldots , [(d-1)/2]\,; \label{PCc}\end{eqnarray}
\begin{eqnarray}
     \big(S(x,p,J)\big)^2-\text{sign}(\varrho)\varrho^2=0\,,\qquad C^L_l(
     S(x,p,J))=\varrho_l\,,\qquad
     l=1,
\ldots , [d/2]-1\,, \label{LCc}\end{eqnarray}  where $S(x,p,J)= J -
x\wedge p$. These constraints on $x,p,J$ are immediate classical
counterparts of the quantum irreducubility conditions (\ref{C-P}),
(\ref{C-L}). Irreducibility of the Poincar\'e group representation
in quantum theory also assumes relation (\ref{C+}) to hold, hence no
other independent constraints, besides (\ref{PCc}), (\ref{LCc}), can
arise at the classical level between position, momentum and angular
momentum. Notice that the ordering ambiguities for the operators
$x,p,J$ cannot affect relations (\ref{PCc}), (\ref{LCc}) because in
the classical limit the distinctions disappear between different
symbols of operators.

Constraints (\ref{PCc}) fix the level of the classical Casimir
functions of the Poincar\'e group, restricting the admissible values
of classical conserved quantities of the particle -- momentum $p$
and angular momentum $J$. These constraints mean that $p$, $J$ are
reduced to the co-orbit of the Poincar\'e group.  Once the
Poincar\'e group representation is irreducible at the quantum level,
any classical path $x(\tau)$ of the particle in Minkowski space has
to satisfy algebraic equations (\ref{LCc}) involving arbitrary
constants of motion $p,J$ subject to constraints (\ref{PCc}) and the
fixed constants $m,s,\varrho$. The algebraic equations (\ref{LCc})
define a surface in Minkowski space. As we explain below in this
section, in any dimension, these surfaces have a timelike axis of
symmetry defined by the particle momentum. The same axis has two
directions distinguished by the sign of $p^0$. We choose $p^0>0$.
Similar ambiguity concerns the sign of proper time, so we chose it
in a consistent way with the sign of $p^0$, i.e. we consider the
causal paths defined by the condition $\dot{x}{}^0>0$.

Once the representation (\ref{hatp}), (\ref{J-S}) is irreducible, it
should completely define the quantum dynamics. Therefore, the
classical dynamics have to be completely defined by the classical
limit of irreducibility conditions. In particular, this means that
the classical particle paths in Minkowski space are defined by the
fact that they satisfy the algebraic equations (\ref{LCc}) with
constant parameters $p,J$ constrained by (\ref{PCc}). Given the
constants of motion -- momentum and angular momentum -- no
requirement can be imposed on the classical path with
$\dot{x}{}^0>0$ other than to lie on the surface defined by
equations (\ref{LCc}). That is why, we refer to the surfaces
(\ref{LCc}) as \emph{world sheets} of the spinning particle.

This understanding of irreducible spinning particle classical
dynamics has consequences. In particular, any two trajectories with
$\dot{x}{}^0>0$  have to be considered as equivalent once they lie
on the same world sheet (i.e., on the surface  with the same
specific values of $p$ and $J$). This understanding may seem quite
different from the usual formulation of classical theory where the
trajectories are defined as solutions of equations of motion, being
an ODE system. The equivalence between different ODE solutions is
usually understood as a consequence of gauge symmetry of the EoMs,
while the constants of motion are the integrals of the ODE system.
In the next section, we demonstrate that the usual form of classical
dynamics can be deduced for the spinning particle from the algebraic
equations of world sheets (\ref{PCc}), (\ref{LCc}).

Now, let us discuss the geometry of the world sheets. Given momentum
and angular momentum subject to classical irreducibility relations
(\ref{PCc}), particle's world sheet is a level surface of the
Lorentz group Casimir functions (\ref{LCc}) in any dimension. At
first, consider the simplest case $d=3$. In $d=3$, Eqs. (\ref{PCc})
read
\begin{equation}\label{PC3}
p^2+m^2=0\,,\qquad (p,J)=ms\,.
\end{equation}
The Lorentz group has a single Casimir function in $d=3$. The level
surface is defined as
\begin{equation}\label{s2d3}
S^2=\mathrm{sign}(\varrho)\varrho^2\, , \qquad
\varrho=\mathrm{const}\,.
\end{equation}
If $\rho <0$, the vector of spin momentum
$S^a=\frac12\epsilon^{abc}S_{bc}$ belongs to a two-sheet
hyperboloid; for $\varrho >0$, $S$ is in the one-sheet hyperboloid.
For $\varrho=0$, the vector of spin lies on the cone.

The compatibility conditions for Eqs. (\ref{PC3}) and (\ref{s2d3})
impose some restrictions on the parameters $\varrho$ and $s$. Using
the second relation (\ref{PC3}), we get the spin constraint,
\begin{equation}\label{pS3}
(p,S)=ms\,.
\end{equation}
Applying the obvious identity
\begin{equation}\label{vpS}
[p,S]^2\equiv(p,S)^2-p^2S^2\geq0\,,
\end{equation}
which is valid for any vector $S$ and timelike vector $p$, we see
that the spin constraints (\ref{s2d3}) and (\ref{pS3}) are
consistent only if
\begin{equation}\label{rS}
s^2+\text{sign}(\varrho)\varrho^2\geq0\,.
\end{equation}
The admissible configurations of spin momentum $S^a$ are defined by
the intersection of the quadric (\ref{s2d3}) and the plane
(\ref{pS3}). In general, it is a circle $S^1$, whose radius is given
by the square root of the lhs of inequality (\ref{rS}),
$r=\sqrt{s^2+\text{sign}(\varrho)\varrho^2}$. The circle can
collapse to a point if $s+\varrho=0$. In this case case, the spin is
constrained to be proportional to momentum
\begin{equation}\label{r0}
S=-\frac{s}{m}p\, .
\end{equation}
We skip this degenerate case where the spin does not have
independent degree of freedom.

Substituting the spin vector $S$ as a function of the conserved
quantities $p$ and $J$ from relation (\ref{SpJx}) into (\ref{s2d3}),
we get the equation of a four-parameter variety of surfaces in $3d$
Minkowski space,
\begin{equation}\label{WSd3}
(J-[x,p])^2=\text{sign} (\varrho)\varrho^2\, .
\end{equation}
The surfaces are parametrized by two conserved vectors $p$ and $J$
subject to two constraints (\ref{PC3}). In the other words, every
element of the variety of $2d$-surfaces (\ref{WSd3}) in Minkowski
space corresponds to the point of the Poincar\'e group co-orbit.

Let us see that the quadric surfaces defined  by equation
(\ref{WSd3}) are cylinders. It is convenient to introduce instead of
$p$ and $J$ two other constant vectors, $n$ and $y$,
\begin{equation}\label{Jy}
 y =\frac{1}{m^2}\left[p,J\right]\,, \qquad n=\frac{p}{m}
\Leftrightarrow  J=m[y,n]- s n\,,\qquad p= m n \, .
\end{equation}
Whenever $p$ and $J$ are constrained by relations (\ref{PC3}), the
vectors $n,y$ are orthogonal to each other and $n$ is normalized,
\begin{equation}\label{ny}
    n^2=-1\, , \qquad (n,y)=0 \, .
\end{equation}
In terms of normalized constant vectors $n,y$ the equation
(\ref{WSd3}) reads
\begin{equation}\label{cylinder3}
(x-y)^2+(n,x)^2=r^2\,, \qquad
r=\frac{1}{m}\sqrt{s^2+\text{sign}(\varrho)\varrho^2}\,.
\end{equation}
It is the equation of a circular cylinder of radius $r$ in $3d$
Minkowski space. The timelike unit vector $n$ is directed along the
axis of cylinder. The vector $y$, being orthogonal to $n$, connects
the origin of reference system with the axis of cylinder. Once $n$
is timelike, $y$ is spacelike. By Eq. (\ref{cylinder3}), the vector
$n=p/m$   defines the direction of the cylinder axis, while the
vector $y$ specifies the position of the axis in space. A cylinder
of fixed radius  with any position of the timelike axis is
admissible. In this way, $n$ and $y$ parameterize the variety of all
possible cylinders with timelike axis and fixed radius.

The case of $s=\varrho=0$ obviously corresponds to a spinless point
particle, so the paths cannot be anything but the straight lines.
With nonvanishing spin, $s\neq 0$, the particle paths should be
cylindric lines.\footnote{In the special case of $\varrho=-s$, the
cylinder collapses to a straight line. We do not address this
degenerate case in this paper. Also notice that in many of the
minimal spinning particle models in $d=4$
\cite{KuLS1,Starus,Bratek1,Bratek2,Bratek3,DasGhosh,Kassaetall,Rempel},
the spin $S_{ab}$ squares to zero, so $\varrho=0$. The radius of
cylinder in this case is just $s/m$.} The latter fact, as we see in
the next sections, is sufficient to completely define the classical
dynamics, including the equations of motion for the particle in
$d=3$.

A similar picture can be seen in four dimensions. Eqs. (\ref{PCc})
restrict admissible values of the conserved quantities -- momentum
and angular momentum -- to the Poincar\'e group co-orbit. There are
two Casimir functions of the Lorentz group. Hence two equations
(\ref{LCc}) are imposed on the particle position in Minkowski space,
given any concrete values of $p$ and $J$. One of the equations for
$x$ is a quadric, and another one defines a plane. The intersection
of the quadric and the plane is a $2d$ cylinder whose timelike axis
is defined by particle's momentum. The radius of cylinder vanishes
in the spinless limit, and the surface degenerates into a timelike
straight line directed along the momentum. Somewhat special is the
case when the plane is tangent to the quadrics, not transverse. We
do not elaborate on this degenerate case here, though it seems
important. \footnote{We plan to address this case elsewhere. We
expect, it corresponds to the irreducible massive spinning particle,
as well as the non-degenerate case. However, the model will have one
gauge symmetry in the degenerate case, in distinction from the pair
of gauge symmetries in general case. The similar special case with
decreased gauge symmetry is known in the minimal model of the
irreducible spinning particle in $d=4$ \cite{LSS}. It is the special
case which admits  consistent inclusion of particle's interaction
with general background fields. }

In $d>4$, the Lorentz group has $[d/2]$ Casimir functions. The
number and order of Eqs. (\ref{LCc}) is growing with dimension, so
the direct study of the system becomes more cumbersome than in lower
dimensions. Instead of direct study of the equations, we use the
reasons of symmetry to establish the structure of particle's world
sheets. At first, consider Eqs. (\ref{LCc}) in the rest reference
system of the particle. Since the equations are covariant, the
conclusions can be extended to a general system by a Poincar\'e
transformation. In the rest system, the momentum and angular
momentum read
\begin{equation}\label{p0J0}
    p=(m,0,\ldots,0)\,,\qquad J=(J^{ij})\,,\qquad
    i,j=1,2,\ldots,d-1\,.
\end{equation}
The condition that the Lorentz boost $J_{0i}$ vanishes is not a
restriction. If $J_{0i}\neq$0, the boost can be absorbed by the
coordinate transformation
\begin{equation}\label{xJp}
x^a\quad\mapsto\quad x^a+\frac{1}{m^2}J^{0a}\,.
\end{equation}
Once $p$ and $J$ are fixed, the group of symmetries of the particle
world sheet equations (\ref{LCc}) is the stabilizer subgroup of the
Poincar\'e group with respect to $p$ and $J$. It includes the
translations along the time axis and spatial rotations that preserve
$J$,
\begin{equation}\label{delx}
x^0 \quad\mapsto \quad x^0+\varepsilon\,,\qquad x^i\quad\mapsto
\quad\Lambda^{i}{}_{j}x^j\,.
\end{equation}
Here, $\varepsilon$ is some constant, and $\Lambda$ is the element
of the stabilizer subgroup $H_J\subset SO(d-1)$ with respect to $J$.
The parameters of transformation (\ref{delx}) are $\varepsilon$ and
$\Lambda$. In general, the transformations (\ref{delx}) are
independent, and their number coincides with the dimension of the
surface defined by equations (\ref{LCc}). Therefore, the world sheet
is the orbit of the symmetry group. The orbit of the translation
subgroup along the fixed axis is a straight line. The stabilizer
subgroup $H_J$ is isomorphic to the maximal torus in $SO(d-1)$
passing through $J$, and it is Abelian. Its orbit is a torus. The
most general $[(d+1)/2]$-dimensional surface, being invariant w.r.t.
both transformations, is a toroidal cylinder
\begin{equation}\label{Csurf}
\Sigma=\mathbb{R}\times \mathbb{T}^D\,,\qquad
\mathbb{T}^D=\underbrace{S^1\times \cdots \times S^1}_{D \text{
times}}\,,\qquad D=[(d-1)/2]\,.
\end{equation}
The levels of the Poincar\'e and Lorentz group Casimir functions
(\ref{C-P}), (\ref{C-L}) determine the radii of the circles
$r_k=r_k(m,s_k,\varrho_l)$. For some special combinations of the
parameters $m,s_k,\varrho_l$ , one or more circles can collapse to
points. The examples of degenerate cases of this sort have been
noticed above in $d=3$ and $d=4$.

As we have observed in this section, any path of irreducible
spinning particle should lie on a (toroidal) cylinder whose position
in the space is defined by particle's momentum and angular momentum,
while the radii are fixed by the representation. We have also seen
that irreducibility of the Poincar\'e group representation does not
impose any other restriction on the world line of the particle. The
latter fact means that any two causal typical world lines have to be
considered equivalent once they belong to the same world sheet.

\section{Equations of motion for world lines on the world sheets}
In this section, we demonstrate that the definition of particle's
paths as causal cylindric lines has consequences. First, it defines
gauge symmetries of the system. Second, it defines the equations of
motion imposed on particle's trajectories. And third, it defines all
the independent conserved quantities $p$ and $J$ as functions of
trajectory and its derivatives. So, the usual formulations of the
classical theory based on equations of motion are deduced from the
fact that equivalent paths form world sheets.

To explain the method we use for deriving the EoMs from the
algebraic equations of particle's world sheet, we discuss at first a
more general problem of the same type we have to solve here.
Consider a manifold $M, \dim{M}=d$ and $m$-parameter set of $n$
smooth functions $C_\alpha (x,y)\in \mathcal{C}^\infty (M) , \alpha
=1, \ldots n$, with $x$ being local coordinates on $M$, and $y_A,
A=1, \ldots , m,$ are the parameters. Consider $m$-parameter variety
of surfaces $\Sigma \subset M$. The element $\Sigma_y$ of the
variety is defined as the zero locus of functions $C_\alpha(x,y)$
with certain parameters $y$,
 \begin{eqnarray}
    \Sigma_y = \{ x \in M \, | \,  C_\alpha (x,y) =0, \, \, \alpha =1, \ldots
    n\} \, ,\qquad \label{Sigma} \dim{\Sigma_y}=d-n\, .
\end{eqnarray}
A simple example of this setup can be the 3-parameter variety of
spheres of fixed radius $r$ in $3d$ Euclidean space. The equation
reads $(x-y)^2-r^2=0$. The components of the radius-vector $y$ of
the sphere center are considered as the parameters.

The line $x(\tau)$ is said to be of $\Sigma$-type if it lies on some
representative of family of surfaces $\Sigma_y$, i.e. $\exists y: \,
x(\tau)\subset \Sigma_y$. Given the set of surfaces $\Sigma_y$, the
problem is to define the class of $\Sigma$-type lines by an ODE
system. For example, the question can be to find the ODE such that
the general solution is the class of spherical lines. The problem of
interest for us is to describe the class of cylindrical lines.

The problem is threefold: (i) to find the ODE system such that the
set of all its solutions coincides with the class of $\Sigma$-lines;
(ii) to define the constant parameters $y$ as integrals of the ODE
system; (iii) to find all the gauge transformations such that they
map a line on $\Sigma_y$ to any other line on the surface with the
same parameters $y$. For example, consider $\Sigma$ being the class
of spheres $S^2$ of fixed radius $r$ in $3d$ Euclidean space. For
the class of spherical lines, the answers are well known to the
first two questions. The ODE of spherical lines (see, e.g.,
\cite{DiffGeom}) reads
\begin{equation}\label{SphODE}
r^2=\frac{1}{\varkappa^2}+\frac{(\varkappa'{})^2}{\omega^2\varkappa^4}\,,
\end{equation}
where $\varkappa$ is the curvature of the line, $\omega$ is the
torsion, and a prime denotes the derivative by the natural parameter
on the line. The parameter $y$, being the radius vector of the
center of sphere, is identified as the integral of motion for the
equation of spherical lines (\ref{SphODE}). It reads
\begin{equation}\label{ySph}
y=x+\frac{1}{\varkappa^2}x''+\frac{\varkappa'}{\varkappa^3\omega}[x',x'']\,,
\end{equation}
The answer to the third question seems unknown in the literature,
though it is quite simple. The infinitesimal gauge symmetry
transformations of the equation of spherical lines (\ref{SphODE})
read
\begin{equation}\label{gsSph}
\delta_{\varepsilon_1}x=[x',[x',x'''\,]]\,\varepsilon_1\,,\qquad
\delta_{\varepsilon_2}x=[x',[x'',x'''\,]]\,\varepsilon_2\,,
\end{equation}
where the infinitesimal transformation parameters
$\varepsilon_1,\varepsilon_2$ are arbitrary functions of parameter
$\tau$ on the curve.

Once the world lines of spinning particles are cylindrical, they
comprise the class of interest in this paper. This class is much
less studied in geometry, though there are some works on cylindrical
lines in $d=3$ Euclidean space \cite{C1,C2}. In Minkowski space, the
issue of general cylindric lines has not been studied yet, to the
best of our knowledge.

Let us explain the general idea of deducing the ODE system for the
lines of $\Sigma$-type, given the system of algebraic equations
which defines the family of surfaces $\Sigma$,
\begin{equation}\label{CSigma}
C_\alpha (x,y) =0\, , \qquad \alpha =1, \ldots n .
\end{equation}
Once the line $x(\tau)$ lies on $\Sigma$, Eqs. (\ref{CSigma}) should
remain valid along the line. This has differential consequences: the
functions $C_\alpha(x(\tau),y)$ have to conserve along the line
$x(\tau)\subset M$ ,
\begin{equation}\label{Cdot}
\dot{C}_\alpha (x,y) \equiv \dot{x}{}^a\frac{\partial {C}_\alpha
(x,y)}{\partial x^a} =0\, , \qquad \alpha =1, \ldots n .
\end{equation}
The same is true for the second and higher derivatives,
\begin{equation}\label{Cddot}
\frac{d^k}{d\tau^k} C_\alpha (x,y)=0 \, , \qquad k=1,2, \ldots \,.
\end{equation}
Under appropriate regularity conditions imposed on $C_\alpha(x,y)$,
all the parameters $y$ can be expressed from Rels. (\ref{Cddot}) as
functions of $x, \dot{x}, \ddot{x},\ldots \, $,
\begin{equation}\label{y-int}
    y_A=\bar{y}_A(x,\dot{x}, \ddot{x}, \dddot{x},\ldots \,) , \qquad
    A=1,2,\ldots,m\,.
\end{equation}
Substituting the functions $\bar{y}_A(x,\dot{x}, \ddot{x},
\dddot{x},\ldots \,)$ instead of all the parameters $y_A$ in the
original  algebraic equations (\ref{CSigma}), we arrive at the ODE
system
\begin{equation}\label{EoM-gen}
C_\alpha (x,\bar{y}(x,\dot{x}, \ddot{x}, \dddot{x},\ldots \,))=0 \,
.
\end{equation}
The functions $\bar{y}_A(x,\dot{x}, \ddot{x}, \dddot{x},\ldots \,)$
are by construction the integrals of motion for this system, so the
solutions lie on the surface $\Sigma_y$ (\ref{Sigma}). In this way,
the ODE system is constructed for the lines of $\Sigma$-type.

Consider the issue of gauge symmetry for the ODE system
(\ref{EoM-gen}). The quantities $\bar{y}$ conserve on solutions,
while the conserved quantities must be gauge invariant. This means
that the gauge symmetry can connect the solutions  which correspond
to the same parameters $y$. It further means that the gauge
transformation should be a symmetry transformation of the surface
$\Sigma_y$ with every fixed set of parameters $y$. This amounts to
saying that the gauge symmetry is generated by the vector fields
tangential to $\Sigma_y$. Let us choose the basis of the vector
field $R_i=R_i^a(x, y)\partial_a, \, i=1,\ldots ,d-n $ in the
tangent bundle of $\Sigma$, i.e.,
\begin{equation}\label{RSigma}
R_i^a(x,y)\left.\frac{\partial C_\alpha (x,y)}{\partial
x^a}\right|_{\Sigma_y} =0 \, , \qquad  T \Sigma_y = \text{span}
\{R_i\}\,.
\end{equation}
Then, the infinitesimal gauge transformation reads
\begin{equation}\label{GTSigma}
    \delta_\epsilon x^a =R_i^a(x, \bar{y}(x,\dot{x}, \ddot{x}, \dddot{x},\ldots
    \,)) \varepsilon^i (\tau ) \, ,
\end{equation}
where the gauge parameters $\varepsilon^i(\tau)$ are arbitrary
functions of $\tau$. By construction, this transformation leaves the
equations (\ref{EoM-gen}) invariant.

Applying this methodology, one can deduce, in principle, the
equations for the lines of various $\Sigma$-types. For spherical
lines this method works very easily leading to the equations of
spherical lines (\ref{SphODE}), their integrals of motion
(\ref{ySph}), and gauge symmetries  (\ref{gsSph}). For general
cylindric lines it also works, in principle, while Eqs.
(\ref{Cddot}) are more complex algebraic relations in the case of
cylinder, so it is much more problematic to explicitly solve them
for the parameters $y$ (\ref{y-int}). In \cite{C1}, a similar
program has been implemented for cylindric lines in $\mathbb{R}^3$.
The differential equation for the lines has been obtained in an
implicit form. In \cite{C2}, another implicit form of the cylindric
line equations has been deduced employing an alternative \emph{ad
hoc} method in $3d$ Euclidean space. To the best of our knowledge,
the problem of cylindric lines has never been solved in Minkowski
space, even in $d=3$. In Minkowski space, however, the class of null
curves is admissible on the cylinders with timelike axis. This does
not have an analogue in the Euclidean case. For this class of
curves, the method of deducing ODE's works very well along the lines
described above. At the first glance, it may seem strange to see the
massive particles propagating by the lightlike world lines. In fact,
this is not so strange because the coordinates are not gauge
invariant quantities for these particles, while the gauge invariants
can have usual properties being typical for massive particles,
including timelike momentum. The higher-derivative models with null
curves were previously known \cite{NerRam1,NerRam2} such that
describe reducible massive representation. In the next subsection,
we deduce the equations for cylindric null curves. These classical
equations of motion correspond to irreducible massive
representation. In Subsection III.B, we deduce the higher-order
equations for timelike cylindrical world lines, though they are
obtained in an implicit form. In Subsection III.C, we introduce an
auxiliary variable, that allows us to obtain the second-order
equations for timelike cylindrical lines in an explicit form. These
equations agree with previously known equations \cite{GKL1}
describing the classical irreducible massive spinning particle.

\subsection{Lightlike world lines on a cylinder with a timelike axis}
 Let $x(\sigma)$ be a null curve parameterized by the pseudoarc length
such that $(x'',x'')=1$, while $x'$ squares to zero. Hereafter in
this section a prime denotes derivative by $\sigma$.

The cylinder of radius $r$ with time like axis is defined by the
equation
\begin{equation}\label{cyl3}
C(x,y,n)\equiv(x-y)^2+(n,x)^2-r^2=0\,,
\end{equation}
where vector parameters $n$ and $y$ satisfy (\ref{ny}). The vectors
$n$ and $y$ parameterize all the possible positions of the cylinder
in space, given the fixed radius $r$. Once the cylinder is a world
sheet of massive spinning particle, the parameters $n,y$ are
connected with particle's momentum and angular momentum by relations
(\ref{Jy}). With two vector parameters subject to two scalar
constraints, the variety of cylinders is a $4$-parameter set of
surfaces in $d=3$ Minkowski space. To express these parameters in
terms of derivatives of the line, one has to consider differential
consequences of the algebraic equation (\ref{cyl3}) up to fourth
order. Equations (\ref{Cddot}) for the cylinder are specified as
 \begin{eqnarray}
(x',x+n(n,x)-y)=0\,,\nonumber\\
(x'',x+n(n,x)-y)+(n,x' )^2=0\,,\nonumber\\
(x''',x+n(n,x)-y)+3(n,x'' )(n,x' )=0\,,\nonumber\\
(x'''',x+n(n,x)-y)+4(n,x''')(n,x')+3(n,x'')^2-1=0\,.
\label{cyl1234}\end{eqnarray}
 Introduce the difference vector $d$
\begin{equation}\label{ddef}
d=x+n(n,x)-y\qquad\Leftrightarrow \qquad y=x+n(n,x)-d \, .
\end{equation}
 The vectors $d$ and $n$ are subject to conditions
\begin{equation}\label{vecd}
(d,n)=0\,,\qquad n^2=-1\,.
\end{equation}
The vector $d$ is normal to the cylinder axis, and it connects the
axis with the current point on the surface. In terms of $d$, the
equation of cylinder (\ref{cyl3}) reads
\begin{equation}\label{dcyl}
d^2=r^2.
\end{equation}
With the difference vector $d$, the differential consequences
(\ref{cyl1234}) are reformulated as
\begin{eqnarray}
(x',d)=0\,,\nonumber\\ (x'',d)+(n,x')^2=0\,,\nonumber\\
(x''',d)+3(n,x'')(n,x')=0\,,\nonumber\\ (x'''',
d)+4(n,x''')(n,x')+3(n,x'')^2-1=0\, .\label{cyld1234}\end{eqnarray}
It is convenient to (\ref{cyld1234}) w.r.t. $n,d$ by decomposing all
the vector quantities in the Frenet-Serret moving frame. The moving
frame has to be adapted to the description of the null curves. It is
given by the TNB triad normalized as
 \begin{eqnarray}
(N,N)=(T,B)=1\,,\qquad (T,T)=(B,B)=(T,N)=(B,N)=0\,.
\label{TNBnull}\end{eqnarray}  The Frenet-Serret formulas have the
form
 \begin{equation}
x'=T\,,\qquad T'=N\,,\qquad N'=-\varkappa T-B\,,\qquad B'=\varkappa
N\,, \label{Frene0}\end{equation}

where the curvature $\varkappa$ reads
\begin{equation}\label{kw0}
\phantom{\sum_a^b}\varkappa=\frac{1}{2}(x''',x''')\,.\phantom{\sum_a^b}
\end{equation}
Decompose the derivatives of $x$ up to the fourth order in the TNB
frame,
 \begin{eqnarray}
x'=T\,,\qquad x''=N\,,\qquad x'''=-\varkappa T-B\,,\qquad
x''''=-\varkappa'T-2\varkappa N\,.\label{DeriveX0}\end{eqnarray}
For the vectors $d$ and $n$ we chose the ansatz
\begin{equation}\label{dnTNB0}
d=\gamma(\beta T-N)\,,\qquad
n=-\frac{\alpha^2\beta^2+1}{2\alpha}T+\alpha\beta N+\alpha B\,,
\end{equation}
which automatically satisfies three equations: (\ref{vecd}) and the
first equation in (\ref{cyld1234}). The system (\ref{vecd}),
(\ref{cyld1234}) has no special solutions with $\alpha=0$, so this
ansatz (\ref{dnTNB0}) is most general.

On substituting (\ref{dnTNB0}) into (\ref{cyld1234}), we get three
non-trivial equations
 \begin{eqnarray}
\gamma-\alpha^2=0\,,\qquad \beta(\gamma-3\alpha^2)=0\,,\qquad
2\varkappa \gamma-4\varkappa\alpha^2+5\alpha^2\beta^2+1=0\,.
\label{d234}\end{eqnarray}  The solution to this system of equations
reads
\begin{equation}\label{sc}
\alpha=(2\varkappa)^{-\frac{1}{2}}\,,\qquad\beta=0\,,\qquad
\gamma=(2\varkappa)^{-1}\,.
\end{equation}
In such a way we arrive at the following equations for the
cylindrical null curves:
\begin{equation}\label{sys03}
2\varkappa\equiv (x''',x''')=\frac{1}{r}\,,\qquad (x',x')=0\,.
\end{equation}
These equations have only one gauge symmetry, being
reparametrization invariance:
\begin{equation}\label{gs0}
\delta_\varepsilon x=x'\varepsilon\,.
\end{equation}
The general recipe of deducing gauge symmetry (\ref{GTSigma}) for
the lines on surface implies a pair of gauge transformations for the
general lines on a cylinder. The null curves are not general lines
because they are subject to an extra differential equation,
$x'{}^2=0$. The extra equation reduces the gauge symmetry.

Relations (\ref{sys03}) can be interpreted as equations of null
helices. Every helix is a cylindrical line, while not every
cylindrical line is a helix. In $d=3$,  any cylindric null curve is
a helix, however, as we see.

In higher dimensions, the null helices have been studied in details
in ref. \cite{Ferrandes}. As is seen from the classification of the
paper \cite{Ferrandes}, the generic classes of null helices with
independent fixed curvatures are defined by more parameters than the
dimension of nondegenerate Poincar\'e group co-orbit. So general
null helices cannot describe an irreducible massive spinning
particle in $d>4$, unlike $d=3,4$. Some null helices in $d>4$ are
not even toroidal cylindric lines. Because of that, we expect that
only some special types of null helices with certain combinations of
curvatures are destined to describe the dynamics of irreducible
massive spinning particles in the dimension greater than four.

Let us find the cylinder parameters $n,y$ as integrals of motion of
equations (\ref{sys03}). Upon the identification $d=-rx''$, the
parameters read,
 \begin{eqnarray}
n=r^{\frac{1}{2}}x'''+r^{-\frac{1}{2}}x'\,,\qquad
y=x+rx'''(x,x''')+x'''(x,x')+x'(x,x''')+r^{-1}x'(x,x')+rx''\,.
\label{ny0}\end{eqnarray}

Let us check that the equations of motion (\ref{sys03}) describe the
irreducible massive spinning particle indeed. The momentum $p$ and
angular momentum $J$ are connected to the cylinder parameters $n,y$
by relations (\ref{Jy}). Substituting (\ref{ny0}) into (\ref{Jy}) we
get $p,J$:
\begin{equation}\label{pJ0}
p=m(r^{\frac{1}{2}}x'''+r^{-\frac{1}{2}}x')\,,\qquad
J=[x+rx'',p]-\frac{s}{m}p\, .
\end{equation}
These quantities conserve on shell by construction. The gauge
invariance of $p,J$ is also obvious. Given the radius-to-spin
relation (\ref{cylinder3}), $p$ and $J$ satisfy the Casimir
constraints (\ref{PC3}) in $d=3$ with mass $m$ and spin $s$. So, the
gauge invariant integrals of motion define the co-orbit of the
Poincar\'e group, which is four dimensional. To prove the
irreducibility of the system, we have to check that the physical
degree of freedom (DoF) number is four, so there are no other
degrees of freedom besides the ones on the co-orbit. The DoF number
can be counted in various ways. Once the equations of motion
(\ref{sys03}) are non-Lagrangian, we count the DoF by using the
formula (8) from Ref. \cite{KLS13}. This counting method does not
appeal to Lagrangian or Hamiltonian formalism. So, the DoF number is
counted for any ODE system by the formula:
\begin{equation}\label{dofno}
    n=\sum_{k=0}k(n_k - r_k -l_k) \,,
\end{equation}
where $n_k$ is the number of the equations of the order $k$ for
every $k$, the $r_k$ is the number of gauge symmetries of order $k$,
$l_k$ is the number of gauge identities of the order $k$. In the
case at hand, these numbers read $n_3=1$ (constant curvature
equation), $n_1=1$ (null-curve condition), $r_0=1$
(reparametrization gauge symmetry). Substituting all the ingredients
into (\ref{dofno}) we get $n=4$, as it should be for a massive
spinning particle in $d=3$.

Notice that in works \cite{NerRam1,NerRam2} the higher-derivative
Lagrangian models are proposed for describing massive spinning
particles by null curves in $d=3,4$. The Lagrangian is a curvature
of the lightlike line in Minkowski space \footnote{This action can
be interpreted as the pseudoarc length, once the velocity squares to
zero.} while no auxiliary internal space is introduced for spinning
degrees of freedom. The equations of motion in these models describe
$10$ DoF in $d=4$ and $6$ in $d=3$, so the theories correspond to
reducible representations. Noether's momentum conserves in these
models and it has a non-vanishing square. These reducible
representations include massive modes, even though the world lines
are lightlike. As noticed in \cite{NerRam1,NerRam2}, the dynamics
can be selected of the irreducible theory by adding the mass-shell
constraint by hand to the variational equations of motion.  In the
view of these previously known facts, it does not seem strange that
the higher-derivative equations for the null curves in Minkowski
space can describe the classical dynamics that corresponds to
irreducible massive representation. The equations of motion
(\ref{sys03}) have been deduced above proceeding from the fact that
all the paths of irreducible spinning particle in $d=3$ have to be
cylindrical lines. As we have seen in this subsection, the
restriction of the class of cylindrical lines to the cylindrical
null curves does not break irreducibility.

\subsection{Timelike world lines on the cylinder with timelike axis}

In the beginning of Section III, the general algorithm was proposed
for deducing the differential equations of motion for the world
lines from the algebraic equations of the world sheet. In the
previous subsection, we have explicitly implemented the algorithm
and deduced the equations of motion with the restriction that the
admissible world lines are null curves. In this subsection, we apply
the same algorithm to deduce the equations for the time like world
lines.

The problem has a subtlety that some cylindric lines do not uniquely
define the cylinder they lie on. In a different wording, the line
can belong to the intersection of two or more cylinders of the same
radius. We consider these lines as atypical. Depending on the
relative position of the two cylinders, the intersection can be
either a closed path or a straight line. In Minkowski space, none of
the closed lines can be a causal path, so these curves are unrelated
to the classical trajectories of spinning particles. We
systematically ignore them. The straight lines constitute a special
set of atypical cylindrical curves which have a smaller gauge
symmetry compared to the typical lines on the same world sheet.
Below in this section (see relation (\ref{gssl})), we demonstrate
that each straight line defines its own class of gauge equivalent
atypical trajectories. These trajectories do not define the world
sheet they lie on. As the physical phase space is isomorphic to the
space of world sheets, the atypical world lines do not represent any
of the equivalent physical evolutions of the particle. For this
reason, they are excluded from the class of physical trajectories.

Let $x(\tau)$ be a timelike curve parameterized by the natural
parameter $\tau$, so we have the normalization condition for the
velocity and its consequences
 \begin{eqnarray}
(x',x')=-1\,,\qquad (x',x'' )=0\,,\qquad
 (x',x''' )+(x'',x'' )=0\,.\label{dxt}
\end{eqnarray}  Throughout this section, prime
denotes derivative by $\tau$.

Differentiating the cylinder equation (\ref{cyl3}) four times and
accounting for identities (\ref{dxt}), we arrive at the relations
connecting the constants $n,y$ with the derivatives of the cylindric
line
 \begin{eqnarray}
(x',x+n(n,x)-y)&&=0\,,\nonumber\\
(x'',x+n(n,x)-y)+(n,x' )^2&&=0\,,\nonumber\\
(x''',x+n(n,x)-y)+3(n,x'' )(n,x' )&&=0\,,\nonumber\\
(x'''',x+n(n,x)-y)+4(n,x''')(n,x')+3(n,x'')^2-(x'')^2&&=0\,.
\label{cyl1234t}\end{eqnarray}  The problem is to solve these
relations w.r.t. the constant parameters $n,y$ expressing them all
in terms of derivatives of the timelike world line $x(\tau )$. We
proceed to the solution of these equations by reformulating them in
terms of the constant vector $n$ and the difference vector $d$
(\ref{ddef}),
\begin{eqnarray}
(x',d)&&=0\,,\nonumber\\ (x'',d)+(n,x')^2&&=0\,,\nonumber\\
(x''',d)+3(n,x')(n,x')&&=0\,,\nonumber\\ (x'''',
d)+4(n,x''')(n,x')+3(n,x'')^2-(x'')^2&&=0\,,\,\,
\label{cyld1234t}\end{eqnarray} where the vectors $d$ and $n$ are
orthogonal to each other, and $n$ is normalized; see (\ref{vecd}).
In terms of $n, d$,  the equation of cylinder (\ref{cyl3}) takes the
form (\ref{dcyl}).

Now, we seek to express  $d$ and $n$ from (\ref{cyld1234t}) in terms
of derivatives of the line. The vector $y$ is defined by Rel.
(\ref{ddef}). Once these vectors are expressed in terms of
derivatives of $x$, the vector-valued functions
\begin{equation}\label{nyint}
n=n(x',x'',x''',x'''')\,,\qquad y=y(x',x'',x''',x''''),
\end{equation}
define the integrals of motion. They conserve whenever the curve
lies on the cylinder with fixed radius $r$. The values of integrals
of motion $n$ and $y$ parametrise all the possible positions of the
cylinder in space. Once the cylinder is a world sheet of massive
spinning particle, the parameters $n,y$ are connected with
particle's momentum and angular momentum by relations (\ref{Jy}).
The ODE describing the timelike curves that lie on the cylinder is
obtained by substituting (\ref{nyint}) into the cylinder equation
(\ref{cyl3}), or equivalently, $d(x',x'',x''',x'''')$ into
(\ref{dcyl}). The equations for $d,n$  (\ref{cyld1234t}) is a system
of polynomial equations, much like Eqs. (\ref{cyld1234}) in the case
of null curves. The system (\ref{cyld1234t}) for the integrals of
timelike curves is more complex, however, comparing to
(\ref{cyld1234}). It has a solution, which we can find only in an
implicit form.

We proceed to solving (\ref{cyld1234t}) as follows. We consider the
overdetermined system consisting of cylinder equation (\ref{dcyl}),
its differential consequences (\ref{cyld1234t}) and constraints
(\ref{vecd}). This system includes seven equations for two vector
unknowns $d$ and $n$, which have six components, given the
constraints between them (\ref{vecd}). At first, we solve the system
consisting of six simplest equations: (\ref{vecd}), (\ref{dcyl}),
and three from (\ref{cyld1234t}). At this stage, the integrals of
motion (\ref{nyint}) are found. Then, we substitute the solution to
the last equation (\ref{cyld1234t}). The consistency condition for
the system gives the ODE for the cylindrical curves. Below, we
elaborate on these manipulations.

It is convenient to rewrite equations (\ref{dcyl}), (\ref{vecd}),
(\ref{cyld1234t}) by decomposing all the vector quantities in the
Frenet-Serret moving frame. The moving frame has to be adapted to
the description of timelike curves. It is given by the TNB triad
normalized as
 \begin{eqnarray}
(T,T)=-1\,,\qquad (N,N)=(B,B)=1\,,\qquad (T,N)=(T,B)=(N,B)=0\,.
\label{TNBt}\end{eqnarray}  The Frenet-Serret differentiation
formulas have the form
\begin{eqnarray}
x'=T\,,\qquad T'=\varkappa N\,,\qquad N'=\varkappa T+\omega
B\,,\qquad B'=-\omega N, \label{Frenet}\end{eqnarray} where the
curvature $\varkappa$ and torsion $\omega$ read
 \begin{equation}\label{kwt}
\varkappa=(x'',x'')^{\frac{1}{2}}\,,\qquad
\omega=\frac{(x',x'',x''')}{(x'',x'')}\,.
\end{equation}
Decompose the derivatives of $x$ up to fourth order in the TNB
frame,
 \begin{equation}\label{DeriveX}\begin{array}{c}\displaystyle
x'=T\,,\qquad x''=\varkappa N\,,\qquad
x'''=\varkappa'N+\varkappa(\varkappa{T}+\omega
B)\,,\\[3mm]\displaystyle
x''''=3\varkappa\varkappa'
T+(\varkappa''+\varkappa^3-\varkappa\omega^2)N+(2\varkappa'\omega+\varkappa\omega')B\,.
\end{array}\end{equation}
Introduce the ansatz for expansion of vectors $d$ and $n$ in the TNB
frame
\begin{equation}\label{dnt}
d=r(\alpha_{1} N+\alpha_{2}\,B)\,,\qquad n=\beta_{1}
T-\beta_{2}(\alpha_{2}N-\alpha_{1}B)\,,
\end{equation}
where the unknown quantities
$\alpha_{1},\alpha_{2},\beta_{1},\beta_{2}$ are subject to the
conditions
\begin{equation}\label{sincos}
\alpha_1^2+\alpha_2^2-1=0\,,\qquad \beta_1^2-\beta_2^2-1=0\,.
\end{equation}
By construction, this ansatz automatically resolves relations
(\ref{vecd}), (\ref{dcyl}) imposed on $d$ and $n$, and the first
equation (\ref{cyld1234t}). The nontrivial equations are

\begin{equation}\label{d234l1}
\varkappa r\alpha_{1}+\beta_{2}^2=0\,,\qquad
3\varkappa\beta_{1}\beta_{2}\alpha_{2}+\varkappa'r\alpha_{1}+\varkappa\omega
r \alpha_{2}=0\,,
\end{equation}

\begin{eqnarray}
((\varkappa''+\varkappa^3-\varkappa\omega^2)\alpha_{1}+
(2\varkappa'\omega+\varkappa\omega')\alpha_{2})r+
4(\varkappa'\alpha_{2}-\varkappa\omega\alpha_{1})\beta_{1}\beta_{2}-
\varkappa^2(3\alpha_{1}^2\beta_{2}^2-7\beta_{2}^2-3)=0\,.
\label{d234l2}\end{eqnarray}  The solution to (\ref{sincos}) and
(\ref{d234l1}) reads
 \begin{eqnarray}
\alpha_{1}=-\frac{1}{r\varkappa}\frac{1}{z^2-1},\qquad
\alpha_{2}=\frac{\varkappa'\,}{\varkappa^2}\frac{1}{3z+r\omega(z^2-1)}\,,\qquad
\beta_{1}=\pm\frac{z}{\sqrt{z^2-1}}\,,\qquad
\beta_{2}=\pm\frac{1}{\sqrt{z^2-1}}\,,
\label{S1C1S2C2}\end{eqnarray}  with $z$ being a root of the
algebraic equation

\begin{eqnarray}
P_1(z)\equiv r^4 \varkappa^4 z^8 \omega^2+6 r^3 \varkappa^4 z^7
\omega-r^2 \varkappa^4 (4 r^2 \omega^2-9) z^6+18 r^3 \varkappa^4 z^5
\omega-r^2 (18 \varkappa^4-6 r^2 \varkappa^4
\omega^2+\varkappa'^2+\varkappa^2 \omega^2) z^4\notag\\+ 6 r
\varkappa^2 \omega (3 r^2 \varkappa^2-1) z^3-(9 \varkappa^2+4 r^4
\varkappa^4 \omega^2-2 r^2 \varkappa^2 \omega^2-9 r^2 \varkappa^4-2
\varkappa'^2 r^2) z^2-6 r \varkappa^2 \omega (r^2 \varkappa^2-1)
z\notag\\-r^2 (-\varkappa'^2-\varkappa^2 \omega^2+r^2 \varkappa^4
\omega^2)=0\,. \label{pol1}\end{eqnarray}
  In terms of the variable $z$ defined by Eq.
(\ref{pol1}), the vectors $d$ and $n$ read
\begin{equation}\label{dvec3}
d=-\frac{1}{\varkappa^2}\frac{1}{z^2-1}x''+\frac{r\varkappa'\,}{\varkappa^3}\frac{1}{3z+r\omega(z^2-1)}[x',x'']\,;
\end{equation}

\begin{eqnarray}
n=\frac{z}{\sqrt{z^2-1}}x'-\frac{\varkappa'\,}{\varkappa^3}
\frac{1}{3z+r\omega(z^2-1)} \frac{1}{\sqrt{z^2-1}}x''-
\frac{1}{r\varkappa^2}\frac{1}{(z^2-1)\sqrt{z^2-1}}[x',x'']\,.
\label{nvec3}\end{eqnarray}  Given $d$ and $n$, the integral of
motion $y$ can be found by formula (\ref{ddef}).

Rels. (\ref{dvec3}), (\ref{nvec3}) determine $n$ and $d$ in an
implicit way, because $z$ is not expressed from Eq. (\ref{pol1})
explicitly. Eqs. (\ref{d234l1}), (\ref{d234l2}) can be approximately
solved if the curvature, torsion and their derivatives are
considered small in comparison to the radius of the cylinder, i.e.,
\begin{equation}\label{1a}
r\varkappa\sim r^2\varkappa'\sim r^3\varkappa''\sim r\omega\sim
r^2\omega'\ll1\,.
\end{equation}
In this approximation, $\alpha_2=\beta_1=1$, while $\alpha_1$,
$\beta_2$ become the small parameters of the same order of magnitude
as the curvature and torsion. It is convenient to express
$\alpha_1$, $\beta_2$ from the second relation (\ref{d234l1}) and
relation (\ref{d234l2}). Leaving only the leading contributions in
the curvature and torsion, we find
 \begin{eqnarray}\label{7a}\notag
\alpha_1=\frac{1}{r}\frac{2r\varkappa\varkappa'\omega+3r\varkappa^2\omega'+9\varkappa^3}{4(\varkappa')^2-3\varkappa\varkappa''}\,,\qquad
\beta_2=r\frac{\varkappa''\varkappa\omega-3\varkappa^2\varkappa'-\varkappa'(2\varkappa'\omega+\varkappa\omega')}{4(\varkappa')^2-3\varkappa\varkappa''}\,.
\end{eqnarray}
Under assumption (\ref{1a}), both these quantities are small, so the
approximation $\alpha_1,\beta_2\ll1$ is consistent. The vectors $d$
and $n$ in this approximation read
 \begin{eqnarray}\label{8a}
n=x'+r\frac{\varkappa''\varkappa\omega-3\varkappa^2\varkappa'-\varkappa'(2\varkappa'\omega+\varkappa\omega')}{4(\varkappa')^2-3\varkappa\varkappa''}x''\,,\qquad
d=\frac{2r\varkappa\varkappa'\omega+3r\varkappa^2\omega'+9\varkappa^3}{4(\varkappa')^2-3\varkappa\varkappa''}x''+r[x',x'']\,.\qquad
\end{eqnarray}
These formulas replace (\ref{dvec3}), (\ref{nvec3}) in the class of
cylindrical curves whose curvature and torsion subject to condition
(\ref{1a}). It is seen that $n$ remains regular once the curvature
tends to zero $\varkappa,\omega\rightarrow0$, while $d$ becomes
singular. The singularity is quite natural because the straight line
does not determine a unique cylinder it belongs to, unlike the
typical cylindric curve.

Substituting the solution (\ref{S1C1S2C2}) to (\ref{d234l2}), we get
another polynomial equation:
 \begin{eqnarray}
P_2(z)\equiv3 \varkappa^4 r^3\omega z^8 +9 \varkappa^4 r^2 z^7-r^3
(-\varkappa^2 \omega^3-2 \varkappa'^2 \omega-\varkappa' \varkappa
\omega'+6 \varkappa^4 \omega+\varkappa \varkappa'' \omega)
z^6\notag\\- r^2 (-7 \varkappa^2 \omega^2-4 \varkappa'^2+3 \varkappa
\varkappa''+9 \varkappa^4) z^5-3 r (r^2 \varkappa' \varkappa
\omega'-4 \varkappa^2 \omega-r^2 \varkappa \varkappa'' \omega+r^2
\varkappa^2 \omega^3+2 r^2 \varkappa'^2 \omega) z^4\notag\\
+r^2 (6 \varkappa \varkappa''-14 \varkappa^2 \omega^2-8
\varkappa'^2-9 \varkappa^4) z^3+3 r (2 r^2 \varkappa'^2 \omega-5
\varkappa^2 \omega+2 r^2 \varkappa^4 \omega+r^2 \varkappa' \varkappa
\omega'+r^2 \varkappa^2 \omega^3-r^2 \varkappa \varkappa'' \omega)
z^2\notag\\+(-9 \varkappa^2+4 \varkappa'^2 r^2+9 r^2 \varkappa^4+7
r^2 \varkappa^2 \omega^2-3 r^2 \varkappa \varkappa'') z-r (r^2
\varkappa^2 \omega^3+2 r^2 \varkappa'^2 \omega+3 r^2 \varkappa^4
\omega-r^2 \varkappa \varkappa'' \omega+r^2 \varkappa' \varkappa
\omega'\notag\\-3 \varkappa^2 \omega)=0\,.
\label{pol2}\end{eqnarray}  For a cylindrical curve the l.h.s. of
(\ref{pol1}) and (\ref{pol2}) must vanish simultaneously.

To keep the contact with the earlier result \cite{C1} on the ODE for
cylindric lines in Euclidean space, we note that the auxiliary
variable $z$ is introduced in the way to get the simplest possible
expressions for $d$ and $n$. This simplification comes with a price
of more cumbersome equations (\ref{pol1}) and (\ref{pol2}) compared
to the Euclidean analogues from Ref. \cite{C1}. In particular, both
the polynomials (\ref{pol1}) and (\ref{pol2}) have eighth order,
while in \cite{C1} the analogues have the orders eight and six. This
difference is insignificant because the orders can always be reduced
by Euclid's algorithm.

Once two of the polynomial equations (\ref{pol1}) and (\ref{pol2})
share a common root, the resultant of polynomials $P_1(z)$ and
$P_2(z)$ must vanish.\footnote{Given two univariate polynomials
$P_1(z)$ and $P_2(z)$ as in (\ref{P1P2}), suppose their roots are
$u_1,\ldots,u_m$ and $w_1,\ldots,w_n$, respectively. The resultant
reads
$$
\text{Res}(P_1,P_2)=a_m^nb_m^n\prod_{i=1}^m\prod_{j=1}^{n}(u_i-w_j)\,.
$$
The resultant vanishes if and only if the polynomials have a common
root.} The resultant can be computed as the determinant of the
Sylvester matrix. Given two polynomials, respectively of degree $m$
and $n$,
 \begin{eqnarray}
P_1(z)=a_0+a_1z+a_2z^2+\ldots+a_m z^m\,,\qquad
P_2(z)=b_0+b_1z+b_2z^2+\ldots+b_nz^n\,, \label{P1P2}\end{eqnarray}
the associated Sylvester matrix $(S_{ij})$ is the $(n+m)\times
(n+m)$ matrix defined as follows:
\begin{equation}\label{SylvMat}
S_{ij}=\left\{%
\begin{array}{ll}
    a_{m+i-j},& i=1,\ldots,n\,,\\& j=i,\ldots,i+m; \\
    b_{i-j},& i=n+1,\ldots,n+m,\\ & j=i-n,\ldots,i-n+m\,,\\
    0,& \text{otherwise}\,.
\end{array}%
\right.
\end{equation}
The pair of polynomial equations (\ref{pol1}), (\ref{pol2}) defines
the cylindric timelike line. These two equations are equivalent to a
single equation  stating that the determinant of the Sylvester
matrix (\ref{SylvMat}) vanishes for the polynomials $P_1(z)$ and
$P_2(z)$ (\ref{pol1}), (\ref{pol2}),
\begin{equation}\label{F0}
\det{ \big(S_{ij}
(r,\varkappa,\omega,\varkappa',\omega',\varkappa'')\big)}=0\,.
\end{equation}
The coefficients of the polynomials, and hence the entries of the
Sylvester  matrix, are the functions of the curvature $ \varkappa$,
torsion $\omega$, their derivatives $\varkappa',\omega',\omega''$,
and radius $r$. This means, Eq. (\ref{F0}) is the fourth-order ODE
for the particle path $x(\tau)$. It is the ODE which defines the
timelike world lines of the particle.

The lhs. of equation (\ref{F0}) is a quite certain quantity being
the resultant of polynomials (\ref{pol1}) and (\ref{pol2})
constructed by the rule (\ref{SylvMat}). The explicit form of the
resultant as a function of curvature, torsion and their derivatives
is too long expression to print out, and it is not very informative.
It seems a matter of principle, however, to establish that the
timelike paths of irreducible massive spinning particle in $d=3$
Minkowski space are defined by a single fourth-order ODE for
$x(\tau)$, without a recourse to any internal space for the spinning
degrees of freedom.

By construction, Eq. (\ref{F0}) should have two gauge symmetries.
Applying the general recipe (\ref{GTSigma}) for constructing the
gauge symmetry to the equation of the cylinder (\ref{dcyl}), we find
the gauge symmetry transformations,
\begin{equation}\label{gscyl3}
\delta_{\varepsilon_1}x=n\,\varepsilon_1\,,\qquad
\delta_{\varepsilon_2}x=[n,d]\,\varepsilon_2\,.
\end{equation}
Here, the vectors $n$ and $d$ are expressed in terms of $x$ and its
derivatives by formulas (\ref{dvec3}), (\ref{nvec3}), and
$\varepsilon_1\,,\varepsilon_2$ are the infinitesimal gauge
transformation parameters being arbitrary functions of $\tau$. In
the limit of small curvature and torsion (\ref{1a}), (\ref{7a}), the
formulas (\ref{gscyl3}) take the form
\begin{equation}\label{gssl}
\delta_{\varepsilon_1}x=x'\,\varepsilon_1\,,\qquad\delta_{\varepsilon_2}x=\frac{1}{\varkappa}x''\,\varepsilon_2\,,
\end{equation}
where $x'$ and $x''$ are the velocity and acceleration vectors in
the natural parametrization of the curve. For the straight lines
$x''=0$, the second gauge transformation (\ref{gssl}) becomes
singular. The remaining gauge symmetry is reparametrization, so each
straight line constitutes its own class of gauge equivalent atypical
curves. By this reason, the straight lines are unconnected by gauge
transformations with the typical cylindrical lines.

The physical DoF number can be counted  for the model (\ref{F0}) by
formula (\ref{dofno}). In the case at hand, we have $n_4=1$ (the
equation (\ref{F0}) of motion involves the fourth-order
derivatives), $r_0=2$ (two gauge symmetry transformations
(\ref{gscyl3}) have the zero order). Substituting all the
ingredients into (\ref{dofno}), we get $n=4$, as it should be for
the massive spinning particle in $d=3$.

The momentum $p$ and angular momentum $J$ are connected to the
cylinder parameters $n,y$ by relations (\ref{Jy}). Substituting
(\ref{dvec3}), (\ref{nvec3}) into (\ref{Jy}) we get $p,J$:

 \begin{eqnarray}
p=&&m\Big(\frac{z}{\sqrt{z^2-1}}x'-\frac{\varkappa'\,}{\varkappa^3}
\frac{1}{3z+\omega_r(z^2-1)}\frac{1}{\sqrt{z^2-1}}x''-
\frac{1}{r\varkappa^2}\frac{1}{(z^2-1)\sqrt{z^2-1}}[x',x'']\Big);\,
\label{pt}\end{eqnarray}
\begin{equation}\label{Jt}
\qquad J=[x-d,p]-\frac{s}{m}p\,,
\end{equation}
with the vector $d$ given by (\ref{dvec3}), and $z$ being the common
root of (\ref{pol1}) and (\ref{pol2}). The quantities (\ref{pt}),
(\ref{Jt}) conserve on shell by construction. Given the
radius-to-spin relation (\ref{cylinder3}), $p$ and $J$ satisfy
constraints (\ref{PC3}) in $d=3$ with mass $m$ and spin $s$. So, the
gauge invariant integrals of motions define the co-orbit of the
Poincar\'e group, which is four dimensional. For the above reasons,
the single fourth-order equation (\ref{F0}) describes  an
irreducible massive spinning particle moving along a timelike path.

\subsection{Explicit second-order equations for the timelike world lines of spinning particle}
In this section, we derive the second-order equations defining the
timelike lines lying on the cylinder with the timelike axis. The
main idea is to introduce an appropriate auxiliary angle-type
variable such that can absorb certain combinations of derivatives of
$x(\tau )$. A similar idea was implemented in the work \cite{C2} for
deducing an explicit form of ODE system for cylindric lines in $d=3$
Euclidean space. We introduce the angle variable in a different way
compared to \cite{C2} that seems us better suited the Minkowski
space specifics and leading to a simpler ODE system. Besides that,
our second order system is variational, while the action functional
has a simple geometric meaning. Furthermore, the same action was
previously known in the minimal spinning particle model suggested in
ref. \cite{GKL1}. In this model, the particle configuration space is
chosen to be $M_4=\mathbb{R}^{1,2}\times S^1$. The factor $S^1$,
however, has been considered just as a fiber over Minkowski space
unrelated to the fact that the particle paths are cylindrical lines
in Minkowski space. Now, we see that the structure of the
configuration space of the minimal spinning particle model in $d=3$
and model Lagrangian follow from the fact that the particle evolves
by the cylindric world sheet in Minkowski space.

In the previous subsection, we have got the fourth-order ODE
(\ref{F0}) defining the cylindrical lines $x(\tau)$. Now, we are
going to diminish the order of the ODE for cylindric lines by making
use of auxiliary angular variable absorbing certain combinations of
derivatives. We start deducing the explicit lower order equations
for cylindric lines by involving the difference vector $d$
(\ref{ddef}) in the ODE formulation. By construction, $d$ is normal
to the cylinder axis, and it connects the axis with the current
particle position on the surface. Once the radius of cylinder is
fixed, the difference vector satisfies the following constraints,
\begin{equation}\label{xd}
   \qquad (n,d)=0\,,\qquad d^2=r^2\,.
\end{equation}
Given the point on the cylinder, the vectors $n$ and $[n,d]$ span
the tangent space to the cylinder. If the curve $x(\tau)$ lies on
the cylinder, its tangent vector $\dot{x}=\frac{dx}{d\tau}$ is
tangential to the cylinder. And vice versa, if the velocity is
tangential to the cylinder all over the line, the path belongs to
the cylinder. Since the cylinder tangent space is spanned by the
vectors $n$ and $[n,d]$, any cylindrical line is defined by the
first-order equations,
\begin{equation}\label{xdedot}
\dot{x}=e_1n+e_2[n,d]\,,
\end{equation}
with $\tau$ being some parameter on the curve $x(\tau)$, not
necessarily natural. The expansion coefficients $e_1$ and $e_2$ can
be arbitrary functions of $\tau$. These coefficients, $e_1$ and
$e_2$, can be understood as einbeins. In Ref. \cite{LSJMP09}, it has
been shown that each einbein induces a gauge transformation of the
ODE system. In the case at hand, the generators of transformations
are the coefficients at einbeins in the ODE system (\ref{xdedot}).
In this way, one can see that the first-order equations
(\ref{xdedot}) automatically have gauge symmetries (\ref{gscyl3}).

Eqs. (\ref{xdedot}) describe particle's path in Minkowski space in
terms of the conserved unit vector $n$ and difference vector $d$.
The latter is subject to constraints (\ref{xd}). Now, we solve the
constraints in a parametric form. The solution depends on the
constants $s$ and $\varrho$ (\ref{PCc}), (\ref{LCc}), which define
the level of the corresponding classical Casimir function. Below, we
mostly consider the case when the spin vector is spacelike or null,
i.e. $\varrho\geq0$. It turns out to be the model with the
configuration space $M_4=\mathbb{R}^{1,2}\times S^1$ considered in
Ref. \cite{GKL1}. The case of the timelike spin can be treated in a
similar way.

The world line is supposed to be timelike and casual, i.e.
\begin{equation}\label{x2x0}
\dot{x}^2<0\,,\qquad \dot{x}{}^0>0\,.
\end{equation}
Provided that $\varrho\geq 0$, the constraints (\ref{xd}) admit a
solution w.r.t. $D$ in terms of the auxiliary null vector $b$:
\begin{equation}\label{db}
d=\frac{s[b,n]-\varrho(b+n(n,b))}{m(n,b)}\, , \qquad b^2=0\,.
\end{equation}
The auxiliary angle variable $\varphi$ is introduced as a
parametrization of $b$, because any null vector is defined (up to
inessential overall factor) by a single angle,
\begin{equation}\label{bphi}
b(\varphi)=(1,-\sin\varphi,\cos\varphi)\,.
\end{equation}
In terms of the auxiliary angle, the spin vector $S$ reads
\begin{equation}\label{Sphi}
S=\frac{sb-\varrho[b,n]}{(b,n)}\,,\qquad S^2=\varrho^2\geq0\,.
\end{equation}
The radius of cylinder is given by
$r=\frac{1}{m}\sqrt{s^2+\varrho^2}\,$.

Let us now rewrite Eqs. (\ref{xdedot}) in terms of the unconstrained
angular variable $\varphi$. Multiplying both sides of (\ref{xdedot})
by $n$ and $[n,d]$, we express the einbeins
\begin{equation}\label{e1e2}
e_1=-(n,\dot{x}{})\,,\qquad e_2=\frac{1}{r^2}([n,d],\dot{x}{})\,.
\end{equation}
Accounting for Eqs. (\ref{ddef}) and (\ref{db}), we get
\begin{equation}\label{e2}
e_1=\frac{(b,\dot{x}{})}{(b,n)}-\frac{s}{m}\frac{\dot{\varphi}{}}{(b,n)}\,,\qquad
e_2=\frac{1}{r^2}([n,d],\dot{d})=\frac{\dot{\varphi}{}}{(b,n)}\,.
\end{equation}
Substituting the vector $b$ (\ref{db}) and the einbeins (\ref{e2})
into (\ref{xdedot}), we arrive at the first-order equations for the
cylindrical curves in the form that involves the auxiliary angular
variable $\varphi$
\begin{equation}\label{xphi}
\dot{x}{}=\frac{(b,\dot{x}{})}{(b,n)}n+
\frac{s}{m}\frac{b}{(b,n)^2}\dot{\varphi}{}-\frac{\varrho}{m}\frac{[n,b]\,\,}{(b,n)^2}\dot{\varphi}{}\,.
\end{equation}
The cylinder parameters $n,y$ are explicitly defined as integrals of
motion for the equations of motion (\ref{xphi}),
 \begin{equation}\label{nint}
n=\frac{\dot{x}{}\Big(1-\displaystyle\frac{2s}{m}\frac{\dot{\varphi}{}}{(b,\dot{x}{})}-\frac{\varrho^2}{m^2}\frac{\dot{\varphi}{}{}^2}{(b,\dot{x}{})^2}\Big)+\dot{x}{}{}^2\Big(
\frac{s}{m}\frac{b\dot{\varphi}{}}{(b,\dot{x}{})^2}+\frac{\varrho^2}{m^2}\frac{b\dot{\varphi}{}{}^2}{(b,\dot{x}{})^3}\Big)}
{\sqrt{-\dot{x}{}{}^2\left(1-\displaystyle\frac{2s}{m}\frac{\dot{\varphi}{}}{(b,\dot{x}{})}-\frac{\varrho^2}{m^2}\frac{\dot{\varphi}{}{}^2}{(b,\dot{x}{})^2}\right)}}
+\frac{\varrho}{m}\frac{[b,\dot{x}{}]}{(b,\dot{x}{})^2}\dot{\varphi}{}\,,
\end{equation}
\begin{equation}\label{yint}
y=\frac{\varrho}{m}\frac{\displaystyle
b\Big(1+\frac{s}{m}\frac{\dot{\varphi}{}}{(b,\dot{x}{})}\Big)}
{\displaystyle(b,\dot{x}{})\sqrt{-\frac{1}{\dot{x}{}{}^2}\Big(1-\frac{2s}{m}\frac{\dot{\varphi}{}}{(b,\dot{x}{})}+
\frac{\varrho^2}{m^2}\frac{\dot{\varphi}{}{}^2}{(b,\dot{x}{})^2}\Big)}}-\frac{s}{m}\frac{[b,\dot{x}{}]}{(b,\dot{x}{})}+\frac{\varrho}{m}n+(x,n)n+x\,.
\end{equation}
  Given the relations between the cylinder parameters
$n,y$ and particle momentum $p$, and angular momentum $J$
(\ref{Jy}), the relations above define the conserved quantities of
the particle in terms of particle's path in the Minkowski space
$x(\tau)$, and auxiliary angle variable $\varphi (\tau)$.
Substituting (\ref{nint}) and (\ref{yint}) into (\ref{Jy}), we get
particle's conserved quantities

\begin{equation}\label{pJint}
p=mn\,,\qquad J=[x,p]-\varrho\frac{[b,\dot{x}{}]}{(b,\dot{x}{})}
+s\frac{\displaystyle\frac{b}{(b,\dot{x}{})}}
{\displaystyle\sqrt{-\frac{1}{\dot{x}{}{}^2}
\Big(1-\frac{2s}{m}\frac{\dot{\varphi}{}}{(b,\dot{x}{})}-
\frac{\varrho^2}{m^2}\frac{\dot{\varphi}{}{}^2}{(b,\dot{x}{})^2}\Big)}}
\,. \end{equation}   Consider the equations for the cylindrical
lines (\ref{xphi}). After excluding the cylinder parameters $n$ and
$y$ by making use of relations (\ref{nint}), (\ref{yint}) and their
consequences (\ref{pJint}), we arrive at the explicit second-order
equations for the particle paths $x(\tau)$ and the auxiliary angular
variable $\varphi(\tau)$,

\begin{eqnarray}
\dot{p}\equiv\frac{d}{d\tau}\left(m\frac{\dot{x}{}\Big(1-\displaystyle\frac{2s}{m}\frac{\dot{\varphi}{}}{(b,\dot{x}{})}-\frac{\varrho^2}{m^2}\frac{\dot{\varphi}{}{}^2}{(b,\dot{x}{})^2}\Big)+\dot{x}{}{}^2\Big(
\frac{s}{m}\frac{b\dot{\varphi}{}}{(b,\dot{x}{})^2}+\frac{\varrho^2}{m^2}\frac{b\dot{\varphi}{}{}^2}{(b,\dot{x}{})^3}\Big)}
{\sqrt{-\dot{x}{}{}^2\left(1-\displaystyle\frac{2s}{m}\frac{\dot{\varphi}{}}{(b,\dot{x}{})}-\frac{\varrho^2}{m^2}\frac{\dot{\varphi}{}{}^2}{(b,\dot{x}{})^2}\right)}}
+\varrho\frac{[b,\dot{x}{}]}{(b,\dot{x}{})^2}\dot{\varphi}{}\right)=0\,,\notag\\
\dot{J}\equiv\frac{d}{d\tau}\left([x,p]+s\frac{\displaystyle\frac{b}{(b,\dot{x}{})}}
{\displaystyle\sqrt{-\frac{1}{\dot{x}{}{}^2}
\Big(1-\frac{2s}{m}\frac{\dot{\varphi}{}}{(b,\dot{x}{})}-
\frac{\varrho^2}{m^2}\frac{\dot{\varphi}{}{}^2}{(b,\dot{x}{})^2}\Big)}}
-\varrho\frac{[b,\dot{x}{}]}{(b,\dot{x}{})}\right)=0\,.
\label{dpJ}\end{eqnarray}
  There are Noether identities between equations
(\ref{dpJ}). One identity is a consequence of relation (\ref{ddef}),
and one more follows from normalization condition $p^2+m^2=0$
identically satisfied by the momentum (\ref{pJint}). The identities
read
\begin{equation}\label{dpJNid}
(p,\dot{p})\equiv0\,,\qquad
\dot{J}+\frac{s}{m}\dot{p}-\frac{1}{m^2}p(J,\dot{p})\equiv0\,.
\end{equation}
Given the identities, the second-order system for the cylindrical
lines involves two independent equations for four variables: three
space-time coordinates $x$ and one auxiliary angular variable
$\varphi$. In particular, it is sufficient to require the normalized
vector $p$ to conserve, while the other equations will follow from
this one.  The gauge symmetries for equations (\ref{dpJ}) read
\begin{equation}\label{dpJgs}
\delta_{\varepsilon_1}x=p\,\varepsilon_1\,,\qquad\delta_{\varepsilon_1}\varphi=0\,;\qquad
\delta_{\varepsilon_2}x=0\,,\qquad\delta_{\varepsilon_2}\varphi=\varepsilon_2\,,
\end{equation} with $\varepsilon_1$, $\varepsilon_2$
being the infinitesimal gauge parameters. The first gauge
transformation is a shift of the world line along the axis of
cylinder. The second gauge symmetry acts by rotations in the plane
orthogonal to the axis. As a particular consequence of the
transformation law (\ref{dpJgs}), the angular variable is a pure
gauge. Given the identities and gauge symmetries, the degree of
freedom number for the system (\ref{dpJ}) can be counted by formula
(\ref{dofno}). It equals to four, as it should be for the
irreducible spinning particle in $d=3$.

By linear combining, Eqs. (\ref{dpJ}) can be brought into the
Lagrangian form with the action functional
 \begin{equation}\label{S}
S[x(\tau),\varphi(\tau)]=\int\left(-m\sqrt{-\dot{x}{}{}^2\Big(1-\frac{2s}{m}\frac{\dot{\varphi}{}}{(b,\dot{x}{})}-\frac{\varrho^2}{m^2}\frac{\dot{\varphi}{}{}^2}{(b,\dot{x}{})^2}\Big)}-\varrho\frac{(\partial_\varphi
b,\dot{x}{})}{(b,\dot{x}{})}\dot{\varphi}{}\right)d\tau\,.
\end{equation}
This action  has been suggested in Ref. \cite{GKL1} to describe the
irreducible spinning particle in $d=3$ by the configuration space
$M=\mathbb{R}^{1,2}\times S^1$, where the factor $S^1$ is the
configuration space of spin. From the viewpoint of the Lagrangian
formalism, the integrals of motion $p, \, J$ (\ref{pJint}) are just
Noether's conserved quantities associated to the Poincar\'e symmetry
of the action.

Let us mention about the geometric interpretation of the action
(\ref{S}). Consider the extremal value of the action
\begin{equation}\label{dl}
S[x(\tau),\varphi(\tau)]\Big|_{\frac{\delta S}{\delta
x}=\frac{\delta S}{\delta\varphi}=0}=-m\int\,(n,\dot{x}{})d\tau\,.
\end{equation}
Obviously, the on-shell value of particle's action is the arclength
of the cylindric line projection on the axis of particle's world
sheet. This provides the interpretation for the particle action from
the viewpoint of geometry of path as such, without appealing to any
fiber bundle over Minkowski space.

\section{Concluding remarks}
Let us briefly summarize what we have observed in this paper about
dynamics of classical massive spinning particles. Once the
Poincar\'e group representation is irreducible and nondegenerate for
quantum spinning particle in Minkowski space, the classical
evolutions of the particle are constrained to the world sheets. In
$d=3,4$ the world sheets are  $2d$ cylinders. In $d>4$ dimensions,
the world sheets are toroidal cylinders $\mathbb{R}\times
\mathbb{T}^D$, with the torus dimension $D=[(d-1)/2]$. The radii of
the cylinders are fixed by representation. Positions of the world
sheets in Minkowski space are defined by particle's conserved
momenta and angular momenta subject to the conditions constraining
them to a nondegenerate co-orbit of the Poincar\'e group. So, the
space of particle's world sheets is isomorphic to the co-orbit of
the Poincar\'e group. All the causal world lines on the same world
sheet are gauge equivalent to each other. The particle paths, being
understood as cylindric lines in Minkowski space, can be defined by
an ODE system. We demonstrate a general scheme of deducing such a
system. The latter ODE system can be understood as classical
equations of motion for the irreducible massive spinning particle.
The equations for the cylindrical lightlike line are explicitly
deduced in $d=3$, and they turn out to be equations of null helices.
Even though the lines are lightlike, the equations describe
classical dynamics of irreducible massive spinning particle. The
timelike cylindrical world lines in $d=3$ are shown to be defined by
a single fourth-order equation with two zero-order gauge symmetries.
This higher-derivative equation of motion is deduced for the
irreducible massive spinning particle in an implicit form. By
introducing an auxiliary angle variable, being a pure gauge degree
of freedom, this equation is shown to reduce to an equivalent
explicit second-order system. The latter equations have been
previously known as EoMs of the minimal model of the $d=3$
irreducible spinning particle \cite{GKL1}.

Overall, we see that the classical dynamics of irreducible spinning
particles in various dimensions are completely defined by their
world sheets in Minkowski space.  The usual form of classical
dynamics, being based on EoMs, is deduced from the world-sheet
formulation. It is also seen that the EoMs of irreducible massive
spinning particle can be formulated in terms of paths in Minkowski
space, without recourse to any internal configuration space
attributed to spinning degrees of freedom.

\begin{acknowledgments} We are very grateful
to A.~Segal, and especially to A.~Sharapov for extensive discussions
on various topics addressed in this work. We thank M.~Bukhtyak,
A.~Mironov, and A.~Penskoi for discussions on and references to the
issues of geometry related to this work. We thank A.~Nersessian for
valuable advice on the literature about spinning particles. We thank
Yu.~Brezhnev and M.~Bukhtyak for their help in computer
manipulations with the systems of polynomial equations.

This work is partially supported by the RFBR grant 16-02-00284 and
by Tomsk State University Competitiveness Improvement Program. S.L.
acknowledges support from the project 3.5204.2017/6.7 of Russian
Ministry of Science and Education.
\end{acknowledgments}


\begin{thebibliography}{11}
\bibitem{Fren} J.~Frenkel.  \emph{Die Elektrodynamik des rotierenden Elektrons},
Z. Phys. \textbf{37},(1926) 243-262.

\bibitem{Fryd} A.~Frydryszak. \emph{Lagrangian Models of the Particles with Spin: The First Seventy
Years}, in: From Field Theory To Quantum Groups, p.151-172 (World
Scientific Publishing , 1996).

\bibitem{Kirillov} A.A. Kirillov, Elements of the theory of group representations
(Springer-Verlag, Berlin, 1976).

\bibitem{Kostant} B.~Kostant,
\emph{Quantization and unitary representations.} Lectures in modern
analysis and applications III (1970) 87-208.

\bibitem{Sour} J.M. Souriau,
\emph{Structure of dynamical systems: a symplectic view of physics.}
Vol. 149. (Springer Science $\&$ Business Media, 2012).

\bibitem{LSS} S.L.~Lyakhovich, A.Yu.~Segal, and A.A.~Sharapov,
\emph{Universal model of a D=4 spinning particle.} Phys. Rev.
\textbf{D 54} (1996) 5223.

\bibitem{KuLS1} S.M.~Kuzenko, S.L.~Lyakhovich, A.Yu.~Segal,
\emph{A geometric model of the arbitrary spin massive particle.}
Int. J . Mod. Phys. \textbf{A 10} (1995) 1529-1552.

\bibitem{KuLSS1} S.M.~Kuzenko, S.L.~Lyakhovich, A.Yu.~Segal,
A.A.~Sharapov. \emph{Massive spinning particle on anti-de Sitter
space.} Int. J . Mod. Phys. \textbf{A 11} (1996) 3307-3329.

\bibitem{Starus} A.~Staruszkiewicz,
\emph{Fundamental Relativistic Rotator}, Acta Phys. Pol. B Proc.
Suppl. \textbf{1} (2008) 109-112.

\bibitem{Bratek1} L.~Bratek,
\emph{Breathing relativistic rotators and fundamental dynamical
systems.} J. Phys. A: Math. and Theor. \textbf{43} (2010) 015208.

\bibitem{Bratek2} L.~Bratek, \emph{Fundamental relativistic rotator: Hessian
singularity and the issue of the minimal interaction with
electromagnetic field.} J. Phys. A: Math. and Theor. \textbf{44}
(2011) 195204.

\bibitem{Bratek3} L.~Bratek, \emph{Spinor particle. An indeterminacy in the motion of
relativistic dynamical systems with separately fixed mass and spin.}
J. of Phys.: Conference Series. \textbf{343} (2012) 012017.

\bibitem{DasGhosh} S.~Das,  S.~Ghosh,
\emph{Relativistic spinning particle in a noncommutative extended
spacetime.} Phys.Rev. \textbf{D 80} (2009) 085009.

\bibitem{Kassaetall}  V.~Kassandrov, N.~Markova,  G.~Schaefer,
A.~Wipf, \emph{On a model of a classical relativistic particle of
constant and universal mass and spin.} J. Phys. A: Math. and Theor.
\textbf{42} (2009) 315204.

\bibitem{Rempel}
T.~Rempel, L.~Freidel, \textit{Bilocal model for the relativistic
spinning particle}, Phys. Rev. \textbf{D 95} (2017) 104014;\,
T.~Rempel and L.~Freidel, \emph{A Classical and Spinorial
Description of the Relativistic Spinning Particle}, \texttt{arXiv:
hep-th/1612.00551}.

\bibitem{LSS6d} S.L.~Lyakhovich, A.A.~Sharapov, K.M.~Shekhter,
\emph{Spinning particle in six dimensions.} J. Math. Phys.,
\textbf{38} (1997) 4086-4103.

\bibitem{LSShd1} S.L.~Lyakhovich, A.A.~Sharapov, K.M.~Shekhter,
\emph{Massive spinning particle in any dimension.(I) Integer spins.}
Nucl. Phys. \textbf{B 537} (1999) 640-652.

\bibitem{LSShd1/2} S.L.~Lyakhovich, A.A.~Sharapov, K.M.~Shekhter,
\emph{A uniform model of the massive spinning particle in any
dimension.} Int. J. Mod. Phys. \textbf{A 15} (2000) 4287-4299.

\bibitem{GKL1}
I.V.~Gorbunov, S.M.~Kuzenko, S.L.~Lyakhovich, \textit{On the Minimal
Model of Anyons},  Int. J. Mod. Phys. \textbf{A 12} (1997)
4199-4215.

\bibitem{AL} K.B.~Alkalaev, S.L.~Lyakhovich,
\emph{On the consistency problem of interactions of (2+ 1) massive
spinning particle.} Mod. Phys. Lett. \textbf{A 14} (1999) 2727-2737.

\bibitem{Pisar} R.D. Pisarski, \emph{Theory of curved paths},
Phys. Rev. \textbf{D 34} (1986) 670-673.

\bibitem{Ply-rigid} M.S. Plyushchay, \emph{Massive relativistic point particle with rigidity},
Int. J. Mod. Phys. \textbf{A 4} (1989) 3851-3865.

\bibitem{Plyu-massless}
M.S. Plyushchay, \emph{Massless Point Particle With Rigidity}, Mod.
Phys. Lett. \textbf{A 4} (1989) 837-847.

\bibitem{RR1} E.~Ramos, J.~Roca,
\emph{W-symmetry and the rigid particle.} Nucl. Phys. \textbf{B 436}
(1995) 529-541.

\bibitem{RR2} E.~Ramos, J.~Roca,
\emph{Extended gauge invariance in geometrical particle models and
the geometry of W-symmetry.} Nucl. Phys. \textbf{B 452} (1995)
705-723.

\bibitem{Ners1} A.~Nersessian, \emph{The Hamiltonian formalism for the generalized rigid
particles}, Theor. Math. Phys. \textbf{117} (1998) 1214-1222.

\bibitem{Ners2} A.~Nersessian, \emph{Lagrangian model of a massless particle on spacelike curves},
Theor. Math. Phys. \textbf{126} (2001) 147-160.

\bibitem{NerRam1} A.~Nersessian, E.~Ramos, \emph{Massive spinning particles and the geometry
of null curves}, Phys. Lett. \textbf{B 445} (1998) 123-128.

\bibitem{NerRam2} A.~Nersessian, E.~Ramos,
\emph{A Geometrical particle model for anyons}, Mod. Phys. Lett.
\textbf{A 14} (1999) 2033-2038.

\bibitem{NerMan} A.~Nersessian , R.~Manvelyan, H.~J.~W.~Muller-Kirsten,
\emph{Particle with torsion on 3-D null curves}, Nucl. Phys. Proc.
Suppl. \textbf{88} (2000) 381-384.

\bibitem{Duval1} C.~Duval, M.~Elbistan, P.~A.~Horvathy, P.-M.~Zhang,
\emph{Wigner-Souriau translations and Lorentz symmetry of chiral
fermions}, Phys. Lett. \textbf{B 742} (2015) 322-326.

\bibitem{Duval2} C.~Duval, P.~A.~Horvathy,
\emph{Chiral fermions as classical massless spinning particles},
Phys. Rev. \textbf{D 91} (2015) 045013.

\bibitem{DiffGeom} Yu.~Aminov, Differential geometry and topology of
curves (CRC Press, 2000).

\bibitem{C1} D.~Keren, E.~Rivlin, I.~Shimshoni, I.~Weiss,
\emph{Reconizing $3d$ objects using tactile sensing and curve
invariants}, J. Math. Imaging and Vision, \textbf{12} (2000) 5-23.

\bibitem{C2} E.~L.~Starostin, G.~H.~M.~van~der~Heijden,
\emph{Characterization of cylindrical curves}, Monatsh. Math.
\textbf{176} (2015) 481-491.

\bibitem{Ferrandes}
A.~Ferrandez, A.~Gimenez and P.~Lucas, \emph{Null helices in
Lorentzian space forms}, Int. J. of Mod. Phys. \textbf{A 16} (2001)
4845-4863.

\bibitem{KLS13} D.~S.~Kaparulin, S.~L.~Lyakhovich, and A.~A.~Sharapov,
\emph{Consistent interactions and involution}, JHEP \textbf{1301}
(2013) 097.

\bibitem{LSJMP09} S.~L.~Lyakhovich, A.~A.~Sharapov, \emph{Normal forms and gauge symmetry of local
dynamics}, J. Math. Phys. \textbf{50} (2009) 083510.

\end{thebibliography}
\end{document}